\documentclass[a4paper]{article}

\usepackage{amsfonts}
\usepackage{amsmath}
\usepackage{amssymb}
\usepackage{graphicx}
\usepackage[dvipdfm]{hyperref}

\def\JOIN{\mathop\vee}

\def\bor{\mathop{\mathord{\lor}\!\!\!\raise4pt\hbox{$\scriptscriptstyle 2$}\,}}
\def\band{\mathop{\mathord{\land}\!\!\!\lower2pt\hbox{$\scriptscriptstyle 2$}\,}}

\newcommand{\CHAIN}[1]{\mathbf{#1}}

\begin{document}
\title{Special Relativity from Causal Ordering}
\title{A Derivation of Special Relativity\\ from Causal Sets}
\author{Kevin H. Knuth\\ Departments of Physics and Informatics\\ University at Albany (SUNY)\\ Albany NY 12222, USA \and Newshaw Bahreyni\\ Department of Physics\\ University at Albany (SUNY)\\ Albany NY 12222, USA }

\date{\today}
\maketitle

\begin{abstract}
We present a derivation of special relativity based on the quantification of causally-ordered events. We postulate that events are fundamental, and that some events have the potential to influence other events, but not vice versa. This leads to the concept of a partially-ordered set of events, which is called a causal set. Quantification proceeds by selecting two chains of coordinated events, each of which represents an observer, and assigning a valuation to each chain. An event can be projected onto each chain by identifying the earliest event on the chain that can be informed about the event. In this way, events can be quantified by a pair of numbers, referred to as a pair, that derives from the valuations on the chains. Pairs can be decomposed into a sum of symmetric and antisymmetric pairs, which correspond to time-like and space-like coordinates. From this pair, we derive a scalar measure and show that this is the Minkowski metric.  The Lorentz transformations follow, as well as the fact that speed is a relevant quantity relating two inertial frames, and that there exists a maximal speed, which is invariant in all inertial frames.  Furthermore, the form of the Lorentz transformation in this picture offers a glimpse into the origin of spin.  All results follow directly from the event postulate and the adopted quantification scheme.
\end{abstract}

\section{Introduction}
In the early 1900s, Albert Einstein radically altered our picture of the universe by doing away with the Newtonian concepts of absolute space and time \cite{Einstein:1905}, and replacing them with relativity and the space-time continuum.  Since then, we have come to imagine space-time to be the fundamental fabric out of which the universe is constructed, yet at the same time we appreciate that different observers can interpret events differently with respect to space and time.  The former perspective provides us with a fundamental framework reminiscent of classical physics, whereas the latter perspective places the observer in a central role similar to what we see in quantum mechanics.

We consider a picture of the universe as being described by a set of events. We do not need to specify precisely what these events refer to, although we visualize them as representing some degree of distinguishability along a chain, which can be used to represent a physical object.  Most importantly, these events do not happen in a space-time.  Instead, the events themselves are considered to be fundamental.  Put simply, events happen.  We assume only minimal additional structure, and assert that some events have the potential to be influenced by other events.  However, this potential is not reciprocal.  That is, if an event $A$ can be influenced by event $B$, then it is not possible that event $B$ can be influenced by event $A$.  The result is that events can be partially ordered.  We stress that we make no assumptions about positions of events in space or time, no assumptions about velocities or angles; we merely assert that some events can be ordered and others cannot.

What follows is a theory describing the physics of events.  That is, we investigate what can be said about events given only their relationships to one another.  The set of events, in conjunction with the ordering relation, gives rise to a partially ordered set.  The theory itself originates from the quantification of the partially ordered set.  This is done by selecting and quantifying a distinguished chain of events called an observer.  By introducing two observers, we can quantify other events in the poset by projecting them onto the two chains and assigning each event with a pair of numbers.  We show that the assigned pairs can be decomposed into a symmetric part, which is related to chains, and an anti-symmetric part, which is related to anti-chains.  The result is a decomposition into one-dimensional time and $n$-dimensional space.  This decomposition naturally results in the \emph{Minkowski metric} as a measure of \emph{distance} between events.  Furthermore, we derive \emph{Lorentz transformations} by considering changes of perspective induced by considering alternate observer chains.  Rather than being fundamental, we find that space-time arises as a construct made to make chains of events look simple.

\section{Partially Ordered Sets of Events}
Our approach relies on the

\begin{quotation}
\noindent\textbf{\emph{Event Postulate}}:~Events are fundamental.  Some events have the potential to be influenced by other events.  However, this potential is not reciprocal.  That is, if event $A$ can be influenced by event $B$, then it is not possible that event $B$ can be influenced by event $A$.
\end{quotation}

This potential to be influenced can be viewed as a binary ordering relation, which relates pairs of events and enables one to impose a partial order.  If event $A$ has the potential to be influenced by event $B$, we say that $A$ \emph{includes} $B$ and write $A \geq B$.  This notion of inclusion is transitive, so that if $A \geq B$ and $B \geq C$, then it is also true that $A \geq C$.  Given any pair of events, it is not necessarily true that one can be informed about the other.  In this case, we say that the events are \emph{incomparable} and write $A || C$.  The relationships $A \geq B$ and $B \geq A$ can only hold simultaneously if $A = B$.

Taken together, a set of events and the described ordering relation results in a partially-ordered set, or \emph{poset}, of events.  Such a poset of events is called a \emph{causal set} \cite{Bombelli-etal-causal-set:1987}.  However, we note that the ordering relation need not assume a strict causal relationship, but only the potential for influence.  Causal sets have been employed in approaches to quantum gravity, and are typically endowed with, or embedded within, a Minkowski geometry exhibiting Lorentz invariance \cite{Bombelli-etal-origin-lorentz:1989}.

We approach the problem from another direction entirely.  Given that the poset is considered to be fundamental, we aim to \emph{derive} a means to quantify events.

Additional structure is introduced to the poset by identifying a distinguished set of events called an \emph{observer}.  An observer is a chain of events, which means that the events are totally ordered so that they occur in succession.  That is, a chain is a set of events $\CHAIN{P}$ such that for all events $x$ and $y$ in $\CHAIN{P}$, we have that either $x \leq y$ or $y \leq x$.  The events describing an observer reflect distinguishable units of change.  Physically, one can imagine them to be generated by a clock.

\begin{figure}[t]
  \begin{center}
  \includegraphics[height=0.30\textheight]{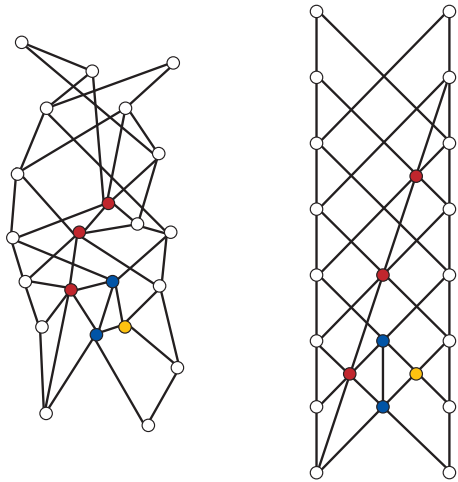}
  \end{center}
  \caption{Both diagrams represent the same poset of events.  In these diagrams, there is no meaning to the horizontal or vertical spacing of the events.  We have the freedom to draw the diagram so as to make chains look simple, but this is merely illustrative.  More importantly, we find that we have the similar freedom to make chains look simple quantitatively. Note that these are not exactly Hasse diagrams as more connections than just the covers are displayed.}
  \label{fig:event-posets}
\end{figure}

As Figure \ref{fig:event-posets} illustrates, depending on how the poset is displayed, chains can be made to look complicated or simple.  The overall goal is to develop a description of events, and we shall do this in such a way to make chains look simple.  Since, at this point, we have no notion of an interval either in space or time, we are at liberty to stretch and squeeze the poset so that certain events in a chain of our choice are drawn at equally-spaced intervals.  While stretching or squeezing a diagram is merely illustrative, we have the similar freedom to make chains look simple quantitatively.

\section{Quantification}
We introduce quantification by assigning a valuation to a chain.  First we select a subset of events on the chain that we will use for quantification.  Not all events on the chain need be used, nor will we display additional events in the subsequent figures.   Events on the chain that are to be used for quantification are assigned a real number such that for any two of these events $x$ and $y$ on the chain $\CHAIN{P}$ related by $x \leq y$ we assign real numbers $p_x \leq p_y$.  We are free to adopt any valuation we please.  To make chains look simple, we assign a valuation, such that for successive quantifying events $x \prec y$, $p_y = p_x + c$ where $c$ is a positive real number.  Without loss of generality, we choose $c = 1$ and label the quantifying events with successive integers.  From now on, we will refer to events on the chain using their label, so that event $p_x$ is assigned a value of $p_x$, where from the context it will be apparent whether $p_x$ refers to the poset element representing the event or its valuation.


An event $x$ can be projected onto a chain $P$ if there exists an event $p \in P$ such that $x \leq p$.  Since any event $p_+ \geq p$ on the chain also includes $x$ by transitivity, and the chain is finite, there must exist a least event $p_x \in P$ such that $p_x \geq x$.  The \emph{projection} of $x$ onto the chain $P$ is given by the least event $p_x$ on the chain $P$ such that $x \leq p_x$.  If one considers the sub-poset consisting only of the element $x$ and the elements comprising the chain $P$, then in this sub-poset $p_x$ covers $x$, $p_x \succ x$.  If the projection exists, the element $x$ can then be ``quantified'' by assigning to the element $x$ the numeric label assigned to the element $p_x \in P$.  Note that this quantification scheme does not ensure that all events in the poset will be quantified.  For example, if the observer cannot be informed about the event, then the event does not project to the chain and thus will not be quantified.

Quantification can be made more rich by introducing another observer.  We implement this by identifying a second chain $\CHAIN{Q}$, and endowing it with a valuation of its own.  Events used for quantification are selected carefully so that they are synchronized to the quantifying events of the first chain $\CHAIN{P}$.  This can be done in the standard way by considering projections of events on $\CHAIN{P}$ onto $\CHAIN{Q}$, and vice versa, and requiring that successive quantifying events on one chain project to successive quantifying events on the other.  Note that a consequence of the synchronization requirement is that not all chains qualify as observers.

\begin{figure}[t]
  \begin{center}
  \includegraphics[height=0.25\textheight]{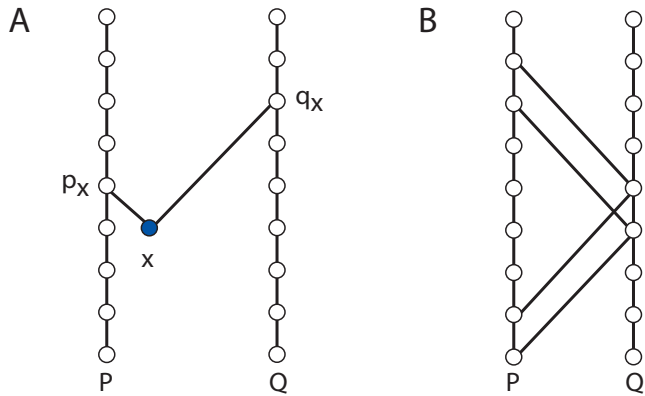}
  \end{center}
  \caption{A) The projection of an event $x$ onto a chain is the minimal event on the chain that can be potentially influenced by $x$, such that for all $p_w \leq p_x$ we have that $x \JOIN p_w = p_x$.  B) Chains can be synchronized by selecting quantifying events on the chains such that successive quantifying events on one chain project to successive quantifying events on the other and vice versa.}
  \label{fig:projection-and-synchronization}
\end{figure}

\subsection{Interval Pair (Pair)}
We want to quantify relationships between events. This requires that we focus on the difference in the way that a \emph{pair of events} is projected onto a pair of chains.  We begin by identifying one way to quantify a pair of events.  In the follow section, we will introduce a second technique and insist that it be consistent with the first.

The first method involves forming a pair of numbers from the direct product of the independent measures obtained by projecting an event onto each of the two reference chains $\CHAIN{P}$ and $\CHAIN{Q}$.  The result is that an event $x$ is quantified by the pair of numbers, $(p_x, q_x)$. To quantify an interval, we designate one event as the origin $0$, and comparing its projection to the projection of another event, which we will label as $1$.  The result is the pair
\begin{equation}
(\Delta p,\Delta q) = (p_1,q_1) - (p_0,q_0) = (p_1-p_0,q_1-q_0).
\end{equation}
From now on, we will suppress the deltas in the notation, and refer to such a pair of differences as an \emph{interval pair}, or more simply as a \emph{pair}.

\begin{figure}[t]
  \begin{center}
  \includegraphics[height=0.40\textheight]{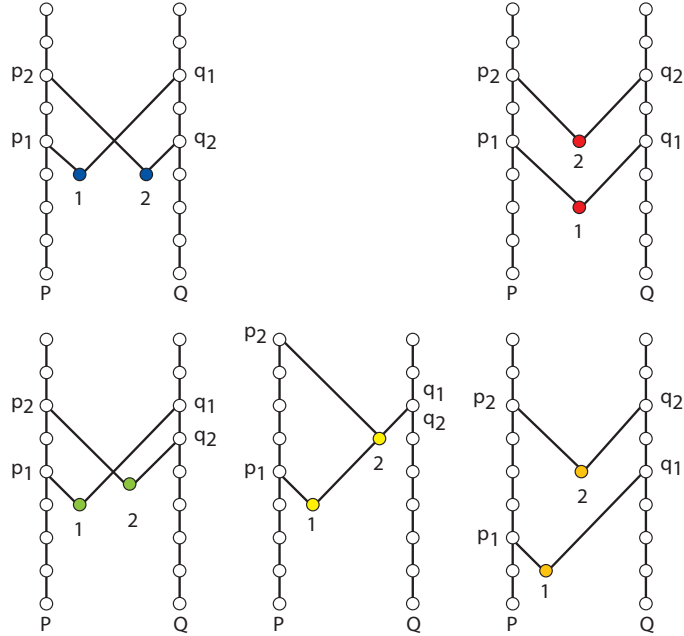}
  \end{center}
  \caption{This figure illustrates five classes of relationships between two events and the observer chains.  (Top Left) These events form an antichain and are recorded in opposite order by the two chains and are interpreted by the observers as being separated only in space. (Top Right) In contrast, these events form a chain and are observed to occur in the same order with respect to the two chains.  They are interpreted by the observers as being separated in time.  (Bottom Left) These events are interpreted as being time-like separated.  (Bottom Center) These events are interpreted as being light-like separated.  (Bottom Right) These events are interpreted as being space-like separated.}
  \label{fig:event-relationships}
\end{figure}

\begin{figure}[t]
  \begin{center}
  \includegraphics[height=0.25\textheight]{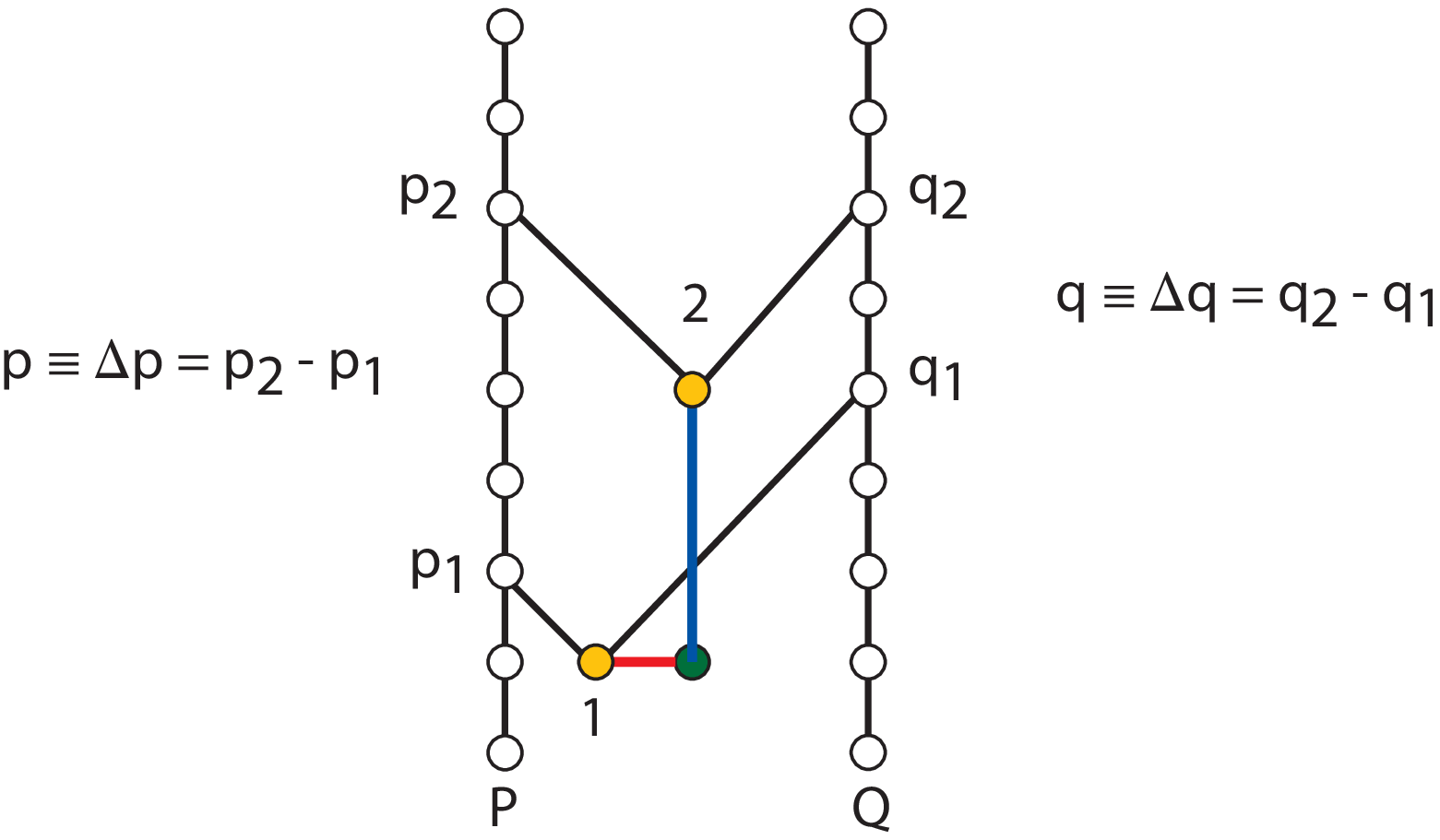}
  \end{center}
  \caption{The interval defined by the events, quantified by the pair of numbers $(p,q)$, can be decomposed into symmetric and antisymmetric parts where the symmetric part is chain-like and the antisymmetric part is antichain-like.}
  \label{fig:decomposition}
\end{figure}

Note that some pairs of events project in such a way that both chains agree as to the order in which they are informed about the events (Figure \ref{fig:event-relationships}, right); whereas other pairs of events project in such a way that the order in which one chain is informed is reverse that of the other chain (Figure \ref{fig:event-relationships}, left).  This fact suggests a convenient decomposition.  Given a pair $(p,q)$, we can decompose it into the sum of a symmetric pair and an antisymmetric pair, such that
\begin{equation}
\label{eqn:decomposition}
(p,q) = \Big(\frac{p+q}{2},\frac{p+q}{2}\Big) + \Big(\frac{p-q}{2},\frac{q-p}{2}\Big).
\end{equation}
 This decomposition, which we call the \emph{symmetric/antisymmetric decomposition}, distinguishes between the two distinct relationships involving differences between pairs of events and the reference chains: that of chains and antichains (Figure \ref{fig:decomposition}).

\subsection{Scalar Measures}
The second method of quantification involves taking the direct product of the chains themselves and forming the unique scalar measure on the product lattice.  Differences are handled by defining the origin of the valuation on each chain to be the minimal projected event of the pair of events.  We aim to identify a unique scalar measure that is a non-trivial function of the pair.  To do this, we define the function $f$ as an unknown map from a pair to a real scalar, and insist that the scalar obeys the symmetric/antisymmetric decomposition (\ref{eqn:decomposition})
\begin{equation}
f(a,b) = f\Big(\frac{(a+b)}{2},\frac{(a+b)}{2}\Big) + f\Big(\frac{(a-b)}{2},\frac{-(a-b)}{2}\Big).
\end{equation}
This functional equation has several solutions:
\begin{eqnarray}
F1.\; f(a,b) & = & a\\
F2.\; f(a,b) & = & b\\
F3.\; f(a,b) & = & ab\\
F4.\; f(a,b) & = & (a + b)^n \;\;\; n \in \;\mbox{odd}\\
F5.\; f(a,b) & = & a^2 + b^2
\end{eqnarray}

We gain some valuable insight by recognizing that this is a special case of the functional equation
\begin{equation} \label{eqn:orthogonality}
f(a_1+b_1,a_2+b_2) = f(a_1,a_2) + f(b_1,b_2).
\end{equation}
where $a = a_1 + b_1$ and $b = a_2 + b_2$ with $a_1 = a_2$ and $b_1 = -b_2$.
We call (\ref{eqn:orthogonality}) the \emph{Orthogonality Relation}, and note that this represents a mapping from a real pair to a real scalar, such that when one adds two pairs, the resulting scalar is also arrived at by simple addition.

Rather than taking the direct product of measures from the two chains and transforming them to a scalar, we can quantify events with a scalar measure assigned to the direct product of the chains.  Consistency requires that the two approaches agree with one another.  The lattice product is associative, which means that the scalar measure also must obey the associativity equation \cite{Aczel:FunctEqns, Knuth:measuring}
\begin{equation} \label{eqn:associativity}
g(f(a,b)) = g(a) + g(b),
\end{equation}
where $g$ is an arbitrary function.

The result is that there are two remaining solutions.  The first solution, $f(a,b) = a+b$, is given by $F4$ with $n=1$ and $g(\cdot)$ equal to the identity.  This solution is proportional to the symmetric component of the decomposition, and is referred to as the \emph{symmetric scalar}.  The symmetric scalar trivially satisfies additivity under the symmetric/antisymmetric decomposition.  However, while the antisymmetric component does satisfy additivity, it does not satisfy associativity and therefore it is not a consistent measure for the interval.  The second solution is $F3$, where we have $f(a,b) = ab$ with $g(\cdot) = log(\cdot)$, so that the scalar associated with the pair $(a,b)$ is $ab$.  We refer to this as the \emph{interval scalar} and denote the interval scalar with the symbol $\Delta s^2$
$$
\Delta s^2 = (p_b - p_a)(q_b - q_a).
$$
keeping in mind that, with respect to the pair, nothing is really being squared. It is straightforward to verify that the interval scalar obeys additivity under this decomposition, since
\begin{equation}
pq = \Big(\frac{p+q}{2}\Big)\Big(\frac{p+q}{2}\Big) + \Big(\frac{p-q}{2}\Big)\Big(\frac{q-p}{2}\Big),
\end{equation}
which can be rewritten as
\begin{equation}
pq = \Big(\frac{p+q}{2}\Big)^2 - \Big(\frac{p-q}{2}\Big)^2.
\end{equation}

Furthermore, it is important to note that any power of the scalar measure can be written in the same form
\begin{equation}
p^k q^k = \Big(\frac{p^k+q^k}{2}\Big)^2 - \Big(\frac{p^k-q^k}{2}\Big)^2,
\end{equation}
where
\begin{equation}
(p^k, q^k) = \Big(\frac{p^k+q^k}{2},\frac{p^k+q^k}{2}\Big) + \Big(\frac{p^k-q^k}{2},\frac{q^k-p^k}{2}\Big).
\end{equation}

\section{Coordinates}
We have shown that the interval pair can be used to form two scalar measures: the interval scalar and the symmetric scalar.  Here we explore the relationships between these scalar measures.  We begin by selecting an event $0$ to serve as the origin.  The interval between event $a$ and the origin is quantified by forming the pair $(p_a-p_0, q_a-q_0)$.  The symmetric scalar, and its antisymmetric counterpart, which is not a proper measure, can be used to define \emph{coordinates} for event $a$ by
\begin{eqnarray}
t_a & = & \frac{(p_a-p_0)+(q_a-q_0)}{2} \nonumber \\
x_a & = & \frac{(p_a-p_0)-(q_a-q_0)}{2}, \nonumber
\end{eqnarray}
so that the pair can be written as $(t_a+x_a,t_a-x_a)$ and decomposed into the sum of two pair
\begin{equation}
(\Delta p, \Delta q) = (p_a-p_0, q_a-q_0) = (t_a+x_a,t_a-x_a) = (t_a,t_a) + (x_a,-x_a),  \nonumber
\end{equation}
each of which depends only on one of the two coordinates.
We can construct similar coordinates for event $b$,
\begin{eqnarray}
t_b & = & \frac{(p_b-p_0)+(q_b-q_0)}{2}  \nonumber \\
x_b & = & \frac{(p_b-p_0)-(q_b-q_0)}{2}, \nonumber
\end{eqnarray}
so that
\begin{equation}
(p_b-p_0, q_b-q_0) = (t_b+x_b,t_b-x_b) = (t_b,t_b) + (x_b,-x_b). \nonumber
\end{equation}
It is easily verified that the interval between events $a$ and $b$ can be quantified by taking the differences of their respective pairs formed with respect to the origin.  The result is that
\begin{equation}
(p_b-p_a, q_b-q_a) = (t_b-t_a,t_b-t_a) + (x_b-x_a,-(x_b-x_a)), \nonumber
\end{equation}
so scalar comparisons can be made simply by constructing the difference between their coordinates.
By denoting such differences in general as
\begin{eqnarray}
\Delta t & = & t_b - t_a  \nonumber \\
\Delta x & = & x_b - x_a, \nonumber
\end{eqnarray}
the pair can be written simply as
\begin{equation}
\label{eq:component-representation}
(\Delta p, \Delta q) = (\Delta t + \Delta x, \Delta t - \Delta x) = (\Delta t,\Delta t) + (\Delta x,-\Delta x).
\end{equation}

Given the coordinate representation of the pair (\ref{eq:component-representation}), we can write the interval scalar as
\begin{equation}
\Delta s^2 = \Delta p \Delta q  = \Delta t^2 - \Delta x^2,
\end{equation}
which we recognize immediately as the Minkowski metric.

\section{Consistency}
\begin{figure}[t]
  \begin{center}
  \includegraphics[height=0.20\textheight]{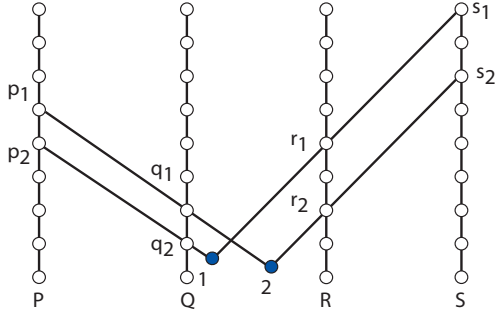}
  \end{center}
  \caption{This figure illustrates two events that are quantified appropriately by chains $\CHAIN{Q}$ and $\CHAIN{R}$, but not by the pair of chains $\CHAIN{P}$ and $\CHAIN{Q}$ or the pair of chains $\CHAIN{R}$ and $\CHAIN{S}$, both of which view the events as simply being distinct in time.}
  \label{fig:unbounded-intervals}
\end{figure}

Given a set of three or more mutually synchronized chains, quantification of an interval via projection to one pair of chains in the set need not necessarily agree with the quantification obtained by projecting to another pair of chains in the same set.  Figure \ref{fig:unbounded-intervals} illustrates an interval that is bounded by some pairs of chains, but not others.  This particular situation is characterized by the fact that some pairs of chains result in a unique time-like quantification, whereas other pairs of chains will result in a consistent quantification.  This time-like quantification can be envisioned by imagining that you and a friend are looking along the same line-of-sight and you each observe two distant flashes, one occurring after another.  In this case it is impossible to determine whether these two flashes originated from the same place at different times, or different places at the same time, or any other situation in between.

Another situation that can occur is one in which the event projects to a set of synchronized chains in such a way that the event is first observed by one chain, followed by its neighboring chains, and so on.  In this case, we must resort to an additional decomposition.

\subsection{The Pythagorean Decomposition}

What follows is a derivation of the Pythagorean theorem, which describes how the space coordinate can be further decomposed into additional coordinates.  Here we derive a decomposition into two spatial coordinates and note that subsequent decompositions into additional spatial dimensions proceed similarly.

\begin{figure}[t]
  \begin{center}
  \includegraphics[height=0.30\textheight]{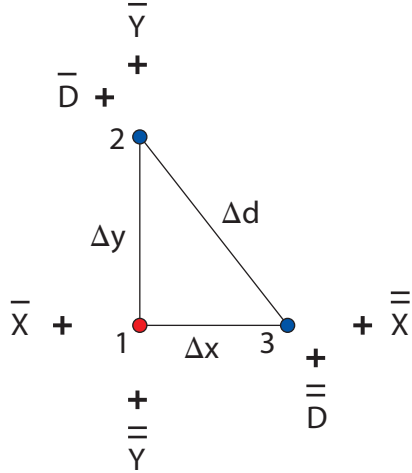}
  \end{center}
  \caption{Here we illustrate how one can decompose the space-like aspect of an interval $\Delta d$ into components $\Delta x$ and $\Delta y$.  This figure represents an `aerial view' of the poset looking down on the reference chains, which are indicated by cross-hairs.  The original interval between events $2$ and $3$ has been `decomposed' by carefully selecting a third event labeled $1$.}
  \label{fig:pythagorean}
\end{figure}

Consider Figure \ref{fig:pythagorean} where a pair of events labeled $2$, $3$ that have been quantified by two chains $\CHAIN{\bar{D}}$ and $\CHAIN{\bar{\bar{D}}}$ and found to have identical time coordinates $t_2 = t_3$, so that the interval pair is $(\bar{\bar{d}}_3-\bar{\bar{d}}_2,\bar{d}_3-\bar{d}_2) = (d_3-d_2,-(d_3-d_2))$, where $\bar{\bar{d}}_i$ represents the projection of event $i$ onto $\CHAIN{\bar{\bar{D}}}$, $\bar{d}_i$ represents the projection of event $i$ onto $\CHAIN{\bar{D}}$, and $d_i$ represents the spatial (antisymmetric) coordinate assigned to event $i$.  We then select a special event $1$ such that its time coordinate is identical to the others $t_1 = t_2 = t_3$.  We introduce chains $\CHAIN{\bar{X}}$ and $\CHAIN{\bar{\bar{X}}}$, which quantify the interval between events $1$ and $3$, and chains $\CHAIN{\bar{Y}}$ and $\CHAIN{\bar{\bar{Y}}}$, which quantify the interval between events $1$ and $2$.  Furthermore, event $1$ is selected so that the three intervals satisfy the orthogonality equation (\ref{eqn:orthogonality}).  We no longer expect the coordinates themselves to sum in a pair-wise fashion since they refer to distinct sets of chains, but we can select an event $1$ so that the scalar interval sums
\begin{equation}
(\bar{\bar{d}}_3-\bar{\bar{d}}_2)(\bar{d}_3-\bar{d}_2) = (\bar{\bar{x}}_3-\bar{\bar{x}}_1)(\bar{x}_3-\bar{x}_1) + (\bar{\bar{y}}_2-\bar{\bar{y}}_1)(\bar{y}_2-\bar{y}_1). \nonumber
\end{equation}
which requires that the $x$ and $y$ coordinates of event $1$ are given by
\begin{eqnarray}
x_1 & = & d_3 - x_3  \nonumber \\
y_1 & = & d_2 - y_2, \nonumber
\end{eqnarray}
since the time coordinates of the three events are equal and events $2$ and $3$ are independent of one another.
Writing the scalar interval in terms of the new coordinates immediately results in
\begin{equation}
(d_3-d_2)^2 = (x_3-x_1)^2 + (y_2-y_1)^2, \nonumber
\end{equation}
which we recognize as the Pythagorean theorem
\begin{equation}
\Delta d^2 = \Delta x^2 + \Delta y^2.
\end{equation}

The result is that the Minkowski metric
\begin{equation}
\Delta s^2 = \Delta t^2 - \Delta d^2, \nonumber
\end{equation}
can be further decomposed into
\begin{equation}
\Delta s^2 = \Delta t^2 - \Delta x^2 - \Delta y^2, \nonumber
\end{equation}
or
\begin{equation}
\Delta s^2 = \Delta t^2 - \Delta x^2  - \Delta y^2 - \Delta z^2, \nonumber
\end{equation}
as necessary.

\subsection{Time and Space}
At this point we have derived that the two classes of coordinates, which we call \emph{space} and \emph{time}, are related to the interval scalar via the Minkowski metric.  Rather than forming a fabric or arena where events take place, space and time are quantifications assigned to make relationships between events and chains of events look simple.  We see that the concepts of time and space arise from the symmetric and antisymmetric decomposition, which originates from the orthogonality of chains and antichains, respectively.  Time, being related to chains, is necessarily one-dimensional.  Space, being related to antichains, can be decomposed into multiple dimensions.

\section{Relating Observers}
\begin{figure}[t]
  \begin{center}
  \includegraphics[height=0.30\textheight]{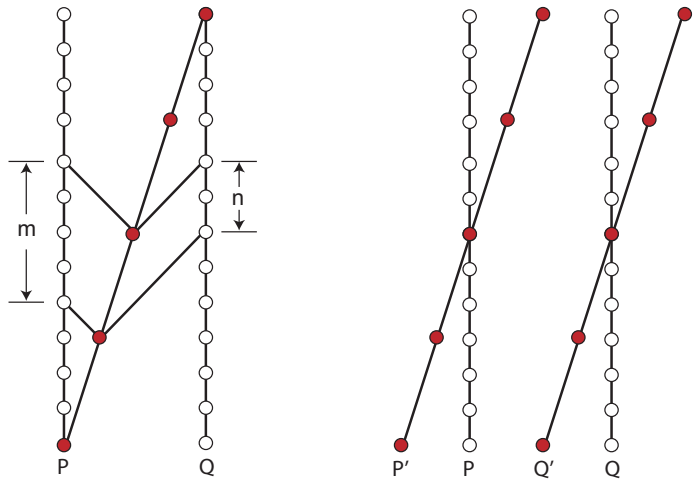}
  \end{center}
  \caption{On the left we illustrate another possible relationship among chains.  The new chain is coordinated to the original pair such that successive events result in projections $\Delta p = m$ and $\Delta q = n$.  This new chain can be used to construct a second pair of observers and thus define another frame of reference.}
  \label{fig:relating-observers}
\end{figure}

We could have chosen another pair of synchronized chains as a basis of quantification.
That is, instead of choosing synchronized chains $\CHAIN{P}$ and $\CHAIN{Q}$, we could have chosen two other synchronized chains $\CHAIN{P'}$ and $\CHAIN{Q'}$, (Figure \ref{fig:relating-observers}) such that successive events in $\CHAIN{P'}$ and $\CHAIN{Q'}$ each comprise an interval which nave non-zero projections $\Delta p = m$ and $\Delta q = n$.  We say that chains $\CHAIN{P}$ and $\CHAIN{P'}$ are \emph{coordinated}, and refer to each pair of chains as an \emph{inertial frame of reference}, or a \emph{frame} for short.

\subsection{Invariance of the Interval Scalar}
We say that the pair of synchronized observers $\CHAIN{P}$ and $\CHAIN{Q}$ comprise frame $1$, and will refer to them from now on as $\CHAIN{P1}$ and $\CHAIN{Q1}$. Similarly, the synchronized pair $\CHAIN{P'}$ and $\CHAIN{Q'}$ comprise frame $2$, where we will refer to them as $\CHAIN{P2}$ and $\CHAIN{Q2}$.  Choosing two events, we denote the interval scalar measured between the events in frame $1$ as $\Delta s_{1}^2$ and the interval scalar measured in frame $2$ as $\Delta s_{2}^2$.  The interval scalar measured in each frame must somehow be related.  More specifically, the interval scalar measured in reference frame $1$ must be a function of the interval scalar measured in reference frame $2$, as well as the only other possible quantities relating the frames, the projections $m$ and $n$.  We also must allow for the fact that observers can choose different scales for labeling their events.  Thus we have the functional relationship
\begin{equation}
\Delta s_{2}^2 = {\sigma_{21}}^2 f(\rho_{12}, \Delta s_{1}^2),
\end{equation}
where $\rho_{21} = \rho(m_{21},n_{21})$ relates the frames by projecting successive events from the chains comprising frame $2$ onto the chains comprising frame $1$, and $\sigma_{21}$ converts the arbitrary scale from one frame to the other.  We can also relate the interval scalar measured in frame $1$ to the interval scalar measured in frame $2$ by
\begin{equation}
\Delta s_{1}^2 = {\sigma_{12}}^2 f(\rho_{12}, \Delta s_{2}^2),
\label{eq:interval-transform}
\end{equation}
where $\rho_{12}$ is a different ratio relating frame $1$ to frame $2$.
Taking the derivative of (\ref{eq:interval-transform}) with respect to $\Delta s_{1}^2$, we find that
\begin{equation}
{\sigma_{12}}^2 {\sigma_{21}}^2 \Big(\frac{df}{d\Delta s_{1}^2}\Big) \Big(\frac{df}{d\Delta s_{2}^2}\Big) = 1. \nonumber
\end{equation}
Since the choice of scale is independent of the transformation, we have that
\begin{equation}
{\sigma_{12}}^2  = \frac{1}{{\sigma_{21}}^2}
\end{equation}
and
\begin{equation}
\Big(\frac{df}{d\Delta s_{1}^2}\Big) \Big(\frac{df}{d\Delta s_{2}^2}\Big) = 1.
\end{equation}

By introducing a third observer and relating three observers to one another in a pairwise fashion, we have
\begin{equation}
\Big(\frac{df}{d\Delta s_{1}^2}\Big) \Big(\frac{df}{d\Delta s_{2}^2}\Big) = \Big(\frac{df}{d\Delta s_{2}^2}\Big) \Big(\frac{df}{d\Delta s_{3}^2}\Big) = \Big(\frac{df}{d\Delta s_{1}^2}\Big) \Big(\frac{df}{d\Delta s_{3}^2}\Big)= 1, \nonumber
\end{equation}
and by comparing pairs of equalities, we find that
\begin{equation}
\Big(\frac{df}{d\Delta s^2}\Big) = \pm 1,
\end{equation}
for any interval $\Delta s^2$.
The second derivative is
\begin{equation}
\Big(\frac{d^2 f}{d\Delta s^2}\Big) = 0,
\end{equation}
which means that the function $f$ is linear in its second argument.  We can then rewrite the original transformation as
\begin{equation}
\Delta s_{2}^2 = {\sigma_{21}}^2 \Delta s_{1}^2 g(\rho_{21}) + C, \nonumber
\end{equation}
where $g$ is a function of $\rho$.
However, we know that in the special case where $\Delta s_{1}^2 = 0$, then it is always true that $\Delta s_{2}^2 = 0$, so $C = 0$ leaving us with
\begin{equation}
\Delta s_{2}^2 = {\sigma_{21}}^2 \Delta s_{1}^2 g(\rho_{21}). \nonumber
\end{equation}
Furthermore, since the first derivative was equal to $\pm 1$, we have that
\begin{equation}
g(\rho) = \pm 1. \nonumber
\end{equation}
However,$g(\rho) = -1$ would change the sign of the interval violating the partial order.  The result is that for all coordinated observers the scalar interval is invariant, up to an arbitrary observer-selected scale,
\begin{equation}
\Delta s_{2}^2 = {\sigma_{21}}^2 \Delta s_{1}^2.
\label{eq:invariance-relation}
\end{equation}

\subsection{Transformations}
We can use the invariance of the interval scalar to relate quantification of an interval in one frame to quantification in another.  Without loss of generality, we can assume that the arbitrary scale is ${\sigma_{12}}^2 = {\sigma_{21}}^2 = 1$.  The projection $p_2$ can be written as some function of $p_1$, so that
\begin{equation}
p_2 = f(p_1, \rho_{21}),  \nonumber
\end{equation}
where $f$ is an unknown function depending on the only possible variables relating the projections: $p_1$ and $\rho_{21}$.  We can write the projection $q_2$ in terms of $q_1$ similarly
\begin{equation}
q_2 = g(q_1, \rho_{21}),  \nonumber
\end{equation}
so that the pair transforms as
\begin{equation}
(p_2, q_2) = (f(p_1, \rho_{21}), g(q_1, \rho_{21})).  \nonumber
\end{equation}

We begin by rewriting the interval scalar (\ref{eq:invariance-relation}) in terms of the projections
\begin{equation}
p_1 q_1 = p_2 q_2 = f(p_1, \rho_{21}) g(q_1, \rho_{21})
\label{eq:invariance-projections}
\end{equation}
where $f$ and $g$ are unknown functions.  Taking the second derivative with respect to $p_1$
\begin{equation}
\frac{d^2(p_1 q_1)}{{dp_1}^2} = 0 = g(q_1, \rho_{21}) \frac{d^2{f(p_1, \rho_{21})}}{{dp_1}^2}   \nonumber
\end{equation}
we find that $f$ must be linear in $p$
\begin{equation}
f(p_1, \rho_{21}) = c p_1 \phi(\rho_{21}) + b,   \nonumber
\end{equation}
where $\phi$ is a function of $\rho$ alone.
In the special case where the frames are synchronized with one another $\rho_{21} = 1$, which implies that $b = 0$
\begin{equation}
f(p_1, \rho_{21}) = c p_1 \phi(\rho_{21}).
\end{equation}
Similarly, by taking the second derivative with respect to $q_1$, we can show that
\begin{equation}
g(q_1, \rho_{21}) = k q_1 \gamma(\rho_{21}).
\end{equation}

By transforming from frame $1$ to frame $2$ and back again, we have
\begin{equation}
p_1 = c^2 p_1 \phi(\rho_{12}) \phi(\rho_{21}).
\end{equation}
This implies that
\begin{equation}
c = \pm 1
\end{equation}
and
\begin{equation}
\phi(\rho_{12}) = \phi(\rho_{21})^{-1}.
\end{equation}
Similarly, considering $q_1$ we have that
\begin{equation}
k = \pm 1
\end{equation}
and
\begin{equation}
\gamma(\rho_{12}) = \gamma(\rho_{21})^{-1}.
\end{equation}
Note that $k$ and $c$ must be of like sign, so that $ck = 1$ and the scalar interval does not change sign.

Rewriting (\ref{eq:invariance-projections})
\begin{equation}
p_1 q_1 = p_1 q_1 \phi(\rho_{21}) \gamma(\rho_{21}),
\end{equation}
which implies that, in general,
\begin{equation}
\phi(\rho) = \gamma^{-1}(\rho).
\label{eq:phi(rho):1}
\end{equation}

Last, we observe that the order in which the chains are represented in the pair is irrelevant, but that interchanging $P$ and $Q$ results in inverting the ratio $\rho$.  This implies that
\begin{equation}
\phi(\rho) = \gamma(\rho^{-1}).
\label{eq:phi(rho):2}
\end{equation}
Equating (\ref{eq:phi(rho):1}) and (\ref{eq:phi(rho):2}) we have that
\begin{equation}
\gamma^{-1}(\rho) = \gamma(\rho^{-1}).
\end{equation}
Taking the derivative with respect to $\rho$, we find that
\begin{equation}
\gamma^{-2}(\rho) = \rho^{-2},
\end{equation}
which implies that
\begin{equation}
\gamma(\rho) = \pm \rho
\end{equation}
and from (\ref{eq:phi(rho):1}) that
\begin{equation}
\phi(\rho) = \pm \rho^{-1},
\end{equation}
where the two functions must be the same sign.

The transformation of pairs of projections is given by the simple relation
\begin{equation}
(p', q') = (p {\rho}^{-1}, q \rho).
\end{equation}
We now determine the function $\rho = \rho(m, n)$ by considering a special case.  Consider an interval defined by two successive events on chain $\CHAIN{Q'}$.  These events have some projection $q'$ onto $\CHAIN{Q'}$, the same projection $p' = q'$ onto $\CHAIN{P'}$ and projections $m$ and $n$ onto chains $\CHAIN{P}$ and $\CHAIN{Q}$, respectively.  This results in the relation
\begin{equation}
(q', q') = (m {\rho}^{-1}(m, n), n \rho(m, n)).
\end{equation}
Equating the elements of the pair on the right side results in
\begin{equation}
m {\rho}^{-1}(m, n) = n \rho(m, n),
\end{equation}
which results in
\begin{equation}
\rho(m, n) = \pm \sqrt{\frac{n}{m}}
\end{equation}
since neither of the projections $m$ nor $n$ are zero.

The final pair transformation from one frame to another is
\begin{equation}
(p', q') = \pm (p \sqrt{\frac{m}{n}}, q \sqrt{\frac{n}{m}}).
\end{equation}
The fundamental nature of the pair of projections is manifest in the simplicity of this transformation.  We observe that this is related to the Bondi k-calculus \cite{Bondi:1980}, and that one can write this relation as a pair-wise multiplication between $({\rho_{21}}^{-1}, \rho_{21})$, which has a scalar measure of unity, and $(p_2, q_2)$ as in Kauffman's iterant algebra \cite{Kauffman:1985}.

Changing variables to the coordinates, mixes the pair resulting in a linear transformation
\begin{equation}
(\Delta t_2 + \Delta x_2, \Delta t_2 - \Delta x_2) = ((\Delta t_1 + \Delta x_1) {\rho_{21}}^{-1}, (\Delta t_1 - \Delta x_1) \rho_{21}),
\end{equation}
which can be represented by a matrix multiplication. Solving for $\Delta t_2$ and $\Delta x_2$, we find that
\begin{eqnarray}
\Delta t_2 & = & \frac{\rho_{21} + {\rho_{21}}^{-1}}{2} \Delta t_1 + \frac{\rho_{21} - {\rho_{21}}^{-1}}{2} \Delta x_1 \\
\Delta x_2 & = & \frac{\rho_{21} - {\rho_{21}}^{-1}}{2} \Delta t_1 + \frac{\rho_{21} + {\rho_{21}}^{-1}}{2} \Delta x_1
\end{eqnarray}

By defining
\begin{eqnarray}
\beta_{21} = \frac{{\rho_{21}}^2-1}{{\rho_{21}}^2+1},
\end{eqnarray}
we find the \emph{Lorentz transformation} in coordinate form
\begin{eqnarray}
\Delta t_2 & = & \frac{1}{\sqrt{1-{\beta_{21}}^2}} {\Delta t_1} + \frac{-{\beta_{21}}}{\sqrt{1-{\beta_{21}}^2}} {\Delta x_1} \\
\Delta x_2 & = & \frac{-{\beta_{21}}}{\sqrt{1-{\beta_{21}}^2}} \Delta t_1 + \frac{1}{\sqrt{1-{\beta_{21}}^2}}  \Delta x_1.
\end{eqnarray}

\section{Speed}
The derivation above reveals that relevant quantity relating two inertial frames is $\beta$, which we recognize as the speed. It is more clearly written in terms of the projections $m$ and $n$ or in terms of the coordinates
\begin{eqnarray}
\beta = \frac{m-n}{m+n} = \frac{\Delta x}{\Delta t}.
\end{eqnarray}
We see that if the two pairs of frames are synchronized with one another, then $m=n=1$ so that $\beta = 0$.  In this case we say that the frames are at rest with respect to one another.  The maximal speed is attained in the limit where $m \rightarrow 0$ or $n \rightarrow 0$, which results in $\beta  \rightarrow  \pm 1$.  Since the interval scalar is given by $ds^2 = \Delta p \Delta q = mn$, we have in this case that $ds^2 = 0$.  Given that the interval scalar is invariant, if $|\beta|$ is unity (maximal) in one frame, it is unity (maximal) in all frames.  We have found that there is an ultimate speed limit that is invariant for all inertial frames.

\section{Conclusion}
We present a picture of the universe where events are fundamental.  By asserting that some events have the potential to be influenced by other events, but that this potential is not reciprocal, we can describe the set of all events as a partially ordered set or poset, which is typically known as a causal set.  Quantification of the poset proceeds by distinguishing particular events on pairs of chains, such that successive quantifying events on one chain project to successive quantifying events on another.  Events can then be labeled by their projections onto the quantifying events of the two reference chains.  This results in a quantification scheme that consists of pairs of numbers, which we show can be mapped to two possible scalar measures; one of which is the interval scalar.  The interval scalar under the symmetric/antisymmetric decomposition gives rise to the Minkowski metric.  The Lorentz transformations are derived, as well as the fact that speed is a relevant quantity that has a maximal value invariant in all inertial frames.  We emphasize that all this is derived without assuming the existence of space or time, motion, constancy of the speed of light, or the principle of relativity.  All follow from the Event Postulate and the adopted quantification scheme.

The ordering of events leads to both the notion of a one-dimensional time and a multi-dimensional space.  Time is distinguished from ordering in the sense that time includes a measure of closeness, which is a result of quantification.  Time is symmetric in the sense that two events separated only in time are observed to occur in the same order by all observers comprising a given inertial frame.  The antisymmetry of space arises from the fact that the order in which two events are observed can be interchanged when considering different observers in the same inertial frame.  In this picture, time is related to chains and space is related to antichains.  The rich mathematics of quaternions and geometric algebra all follow from this simple fact.

In addition to the Event Postulate, we have made an assumption about the poset structure.  Specifically, we assume that events are sufficiently dense so that we are able to construct synchronized chains of events as well as identify events that enable us to decompose intervals into orthogonal components.  While such an assumption leads to special relativity and Minkowski space-time, it may require modification to obtain cosmological expansion or gravity.

The simplicity of the Lorentz transformation when expressed in terms of pairs has been noted before by Bondi \cite{Bondi:1980} and later explored by Kauffman \cite{Kauffman:1985}.  Though while attaining a similar formalism, these authors worked within the usual space-time framework.  We approached the problem with the goal of  consistently quantifying a partially-ordered set of events, and consequently arrive at space-time via a convenient decomposition.  After submission of the first version of this paper to the arXiv, we have been introduced to the work of Giacomo Mauro D'Ariano \cite{DAriano:2010} who showed how the Lorentz transformations can be, in principle, derived from event-counting performed by an observer within a causal network implemented by a quantum computer. D'Ariano's approach is similar in spirit to ours in that causality plays a central role, however it differs in that it lies within a quantum mechanical framework.  We find that the only necessary feature is the causal relationship, which we represent by a partially ordered set of events.

It is surprising that arbitrary powers of the interval scalar can be decomposed into the same Minkowskian form.  The fact that the Lorentz transformation depends on the square roots of the projections rather than the values of the projections themselves suggests that square roots of projections are of fundamental importance.  If one rotates the spatial coordinates by a given angle, the projections rotate by the same amount so that a rotation of $2\pi$ brings the coordinates and the projections back to the initial state.  However, if one instead considers quantities dependent on the square roots of projections, these will rotate at one half the rate so that a rotation of $4\pi$ is necessary to return them to the original state.  This suggests that the square roots of projections are described by the spin group Spin($n$), which is a double-cover of the special orthogonal group SO($n$) of rotations.  This does not affect the Lorentz transformations since the square roots of projections appear as ratios where the effect of any rotation is canceled.  Therefore the fact that these fundamental quantities require that one complete two full rotations rather than a single full rotation does not affect the mechanics and is not readily physically apparent.

It may be of interest to the reader to note that while solution F3 of the Orthogonality Relation (\ref{eqn:orthogonality}) along with the symmetric/antisymmetric decomposition gives rise to the square in the metric, solution F5 gives rise to the square in the Born Rule of quantum mechanics.  More importantly, our recent explorations into the foundations of quantum mechanics postulated that pairs of numbers are required to quantify quantum amplitudes \cite{GKS:PRA}.  Here we find pairs of numbers again playing a critical role in the fundamental formulation of special relativity.  It is expected that these results will provide new insights into the meaning of the pair in quantum mechanics.  It is entirely possible that the pair in quantum mechanics is comprised of square roots of projections rather than the projections themselves.  This would lead to a natural description of Fermions (spin-1/2 particles).  Indeed, we already have recognized connections between the proposed method of poset quantification and the Feynman checkerboard \cite{Feynman&Hibbs} where the right and left moves correspond to projections (or more likely, the square roots of projections) onto the two chains.  Furthermore, the more fundamental understanding of the metric introduced here has the potential to facilitate the union of quantum mechanics and gravity resulting in quantum gravity.  Already in our previous work, we showed that quantum theory is more fundamental than space \cite{GKS:PRA}.  Here we show that space-time itself is not fundamental, but rather is a convenient construct chosen to make events look simple.

Given the belief that physical law reflects an underlying order, one would expect that given this underlying order, one ought to be able to derive the most fundamental aspects of physical law.  This fundamental principle is outlined in an earlier work \cite{Knuth:laws}, and has been recently demonstrated to provide insights into the foundations of quantum mechanics \cite{GKS:PRA}.  Here we apply this methodology to understand the physics of events and derive specific notions of time, space and motion.

\section{Acknowledgements}
Kevin Knuth would like to thank Philip Goyal, Keith Earle, Ariel Caticha, John Skilling, Seth Chaiken, Adom Giffin, Jeff Scargle and Jeffrey Jewell for many insightful discussions and comments.  He would also like to thank Rockne, Ann and Emily Knuth for their support, and Henry Knuth for suggesting that he `try using a J'.  Newshaw Bahreyni would like to thank Shahram Pourmand for his helpful discussions and Mahshid Zahiri, Mohammad Bahreyni and Shima Bahreyni for their continued support.  The authors would also like to thank Giacomo Mauro D'Ariano, Alessandro Tosini, Cristi Stoica and Patrick O'Keefe for valuable comments that have improved the quality of this work.

\bibliographystyle{amsplain}
\bibliography{knuth}

%
%
%
%

\end{document}